# Determination of intrinsic spin Hall angle in Pt

Yi Wang, Praveen Deorani, Xuepeng Qiu, Jae Hyun Kwon, and Hyunsoo Yang[a]

*Department of Electrical and Computer Engineering, National University of Singapore,*

*117576, Singapore*

The spin Hall angle in Pt is evaluated in Pt/NiFe bilayers by spin torque ferromagnetic resonance (ST-FMR) measurements, and is found to increase with increasing the NiFe thickness. To extract the intrinsic spin Hall angle in Pt by estimating the total spin current injected into NiFe from Pt, the NiFe thickness dependent measurements are performed and the spin diffusion in the NiFe layer is taken into account. The intrinsic spin Hall angle of Pt is determined to be 0.068 at room temperature, and is found to be almost constant in the temperature range 13 – 300 K.

[a]Electronic mail: eleyang@nus.edu.sg

Pure spin currents have drawn tremendous attention for the applications in emerging spintronics devices due to its low power consumption and high efficiency.[1-5] The efficient generation of pure spin currents and manipulation of the magnetization of a ferromagnetic layer is of central importance in the field of spintronics. The spin Hall effect (SHE) is one of the promising ways to generate spin currents, in which a charge current can be converted to a transverse spin current due to the spin-orbit interaction.[6,7] This process is described by $J_s = \theta_{sh}(\hat{\sigma} \times J_c)$, where $\hbar J_s/2e$, $J_c$, and $\hat{\sigma}$ are the spin current density, charge current density, and spin moment, respectively. Here, $\theta_{sh} = |J_s|/|J_c|$ is the spin Hall angle, which is a measure of the conversion efficiency between charge currents and pure spin currents, and it is a material dependent parameter.

Recently, there have been significant disagreements in the values of $\theta_{sh}$ in Pt reported by different groups, ranging from 0.012 to 0.12.[8-31] One of the effective techniques used to evaluate $\theta_{sh}$ is spin torque ferromagnetic resonance (ST-FMR) measurement,[9] which is reported as a self-calibration method. However, recent reports have suggested that the spin Hall angle evaluated by ST-FMR measurement depends on the thickness of ferromagnetic layer (FM). This dependence was believed to arise from the inverse spin Hall effect (ISHE) due to the spin pumping at the Pt/FM interface[11,32-34] and/or spin transfer torque induced by radio frequency (RF) current in FM layers.[35] In this work, we attribute a FM thickness dependence to spin diffusion in the FM layer. Thus we carry out FM thickness dependent ST-FMR measurements and evaluate the intrinsic spin Hall angle in Pt by taking spin diffusion in FM into account. Furthermore, so far the variation of $\theta_{sh}$ as a function of temperature has been little studied,[36,37] which is of great importance not only for the fundamental understanding of SHE but also for the device applications. In this Letter, we demonstrate the $Ni_{81}Fe_{19}$ (Py) thickness and temperature dependence of $\theta_{sh}$ in Pt/Py bilayers. The intrinsic $\theta_{sh}$ of Pt is determined to be 0.068 at room temperature by taking the spin diffusion in Py into account. Moreover, it is found that $\theta_{sh}$ remains almost constant as temperature decreases from 300 to 13 K.

As shown in Fig. 1(a), multiple films of Pt ($d = 6$ nm)/ $Ni_{81}Fe_{19}$ ($t = 2 - 10$ nm) were deposited onto a thermally oxidized Si wafer at room temperature by using magnetron sputtering with a base pressure of $3\times10^{-9}$ Torr. Subsequently, the films were patterned into rectangular shaped microstrips with dimensions of $L$ (130 μm) $\times$ $W$ (15 − 20 μm) as denoted in Fig. 1(b). In the next step, coplanar waveguide (CPW) were fabricated. Different gaps (25 − 55 μm) between ground (G) and signal (S) electrodes were designed to tune the device impedance to be ~ 50 Ω. An RF current from a signal generator (SG) was applied to the device via a microwave GSG probe and a dc voltage was detected by a voltmeter, simultaneously. An in-plane external magnetic field $H_{ext}$ at a fixed angle ($\theta_H$) of 25 degree with respect to the microstrip length direction was swept between −4 and +4 kOe, and the measurements were performed with a nominal RF power of 13 dBm at room temperature, unless otherwise specified.

When an RF current is applied to the Pt/Py bilayer, a dc voltage is detected by mixing the RF current and the time-varying anisotropic magnetoresistance, similar to the spin diode effect.[38] The dc voltage could be written as $V_{mix} = SF_{sym}(H_{ext}) + AF_{asym}(H_{ext})$,[9] where $F_{sym}(H_{ext}) = \Delta H^2/[(H_{ext} - H_0)^2 + \Delta H^2]$ is a symmetric Lorentzian function centered at the resonant field ($H_0$) with a linewidth ($\Delta H$), and $F_{asym}(H_{ext}) = \Delta H(H_{ext} - H_0)/[(H_{ext} - H_0)^2 + \Delta H^2]$ is an antisymmetric Lorentzian function. Figure 1(c) shows the measured ST-FMR signals (open symbols) on a Pt(6 nm)/Py(5 nm) device at RF frequencies ($f$) spanning 6 - 12 GHz and corresponding fits (solid lines). From the fitting, the values of $S$ and $A$ were determined. As shown in Fig. 1(d), the data of $f$ as a function of $H_0$ for different Py thicknesses match well with the Kittel formula $f = (\gamma/2\pi)[H_0(H_0 + 4\pi M_{eff})]^{1/2}$, where $\gamma$ is the gyromagnetic ratio. By fitting, the effective demagnetization fields ($4\pi M_{eff}$) for Pt/Py ($t$ nm) devices were determined. The saturation magnetization ($M_s$) of Pt/Py ($t$ nm) devices were measured independently by vibrating sample magnetometer (VSM). Using the obtained $4\pi M_{eff}$ = 0.851 ± 0.05 T and $M_s$ = 6.97×10$^5$ A/m for Pt (6 nm)/Py (5 nm) device, the corresponding spin Hall angle is determined to be ∼ 0.063 using $\theta_{sh} = J_s/J_c = (S/A)(e\mu_0 M_s t d/\hbar)[1 + (4\pi M_{eff}/H_{ext})]^{1/2}$.[9] We have also confirmed this measured spin Hall angle value by separate calibration method from only the symmetric component ($S$) of ST-FMR signals.[39,40]

Similar measurements have been further performed for Pt (6 nm)/Py (5 nm) device at different external magnetic field angles ($\theta_H$) and $f$ = 8 GHz. The normalized values of $S$ (symmetric component) and $A$ (antisymmetric component) extracted from the ST-FMR fitting as a function of $\theta_H$ were plotted in Fig. 2(a). Both can be fitted by $cos^2(\theta_H)\,sin(\theta_H)$ denoted by the solid lines, which is consistent with the equation describing the ST-FMR signal.[9] Note that the largest ST-FMR signal is present at an angle of $\theta_H$ = 35° (rather than 45°).[39] In Fig. 2(b), the measured spin Hall angle ($\theta_m$) is observed to be independent on $\theta_H$. We also confirm that the response of Pt (6 nm)/Py (5 nm) device in our measurements is in the linear regime with increasing the power.[39]

Similar ST-FMR measurements have been carried out for devices with different Py thicknesses, and three devices were measured for each Py thickness. The *measured* spin Hall angle ($\theta_m$) in Pt/Py ($t$) devices as a function of Py thickness are obtained from the ratio of $S/A$ as well as by separate calibration method from only the symmetric component ($S$) of ST-FMR signals.[39,40] We did not find a significant difference between results from these two methods.

The Py thickness dependence of *measured* spin Hall angle ($\theta_m$) by separate calibration method is presented in Fig. 3(a). In principle, the spin Hall angle is an intrinsic property of Pt and it should remain constant in various Pt/Py($t$) devices. Instead, the $\theta_m$ is found to increase monotonically (but not linearly) and eventually saturate as the Py thickness increases from 2 to 10 nm. The previously reported zero-thickness interpolation method[11] to get the *intrinsic* $\theta_{sh}$ seems dubious in our case, because $\theta_m$ varies non-linearly with $t$, and hence the zero thickness limit does not lead to an unique value [see a, b and c lines in Fig. 3(a)].

In fact, the total torque exerted by the spin current on the Py layer increases with the Py thickness and saturates in the limit of large thickness of Py.[41] At smaller thicknesses of Py, the spin Hall angle is obviously underestimated. Therefore, we propose that the spin diffusion in Py must be taken into account to reappraise the total spin current injected into Py and thus the spin Hall angle. For our Pt/Py($t$) spin Hall system, the total torque exerted by spin current on Py per unit interface area can be written as[41]

$\tau_{ST}^{total} = J_s \, cos\,(\theta_H)(k_0^2/k^2)[\,cosh\,(kt) - 1]/\,cosh\,(kt)$ , here $k = \sqrt{\lambda_S^{-2} + \lambda_\phi^{-2}}$ , $k_0 = \sqrt{\lambda_\phi^{-2}}$ and $J_s = \theta_{sh} J_e$, where $\lambda_\phi$ is the spin decoherence length in Py, $\lambda_S$ is the spin diffusion length in Py, $J_s$ is the spin current density at the Pt/Py interface, and $J_c$ is the charge current density in Pt layer. For Py with a thickness much larger than the characteristic length $1/k$, the torque saturates and equals $\theta_{sh}J_c\,cos(\theta_H)k_0^2/k^2$ (denoted as $\tau_{ST}^\infty$ thereafter). Hence the total torque is $\tau_{ST}^{total}=\tau_{ST}^\infty\dfrac{cosh(kt)-1}{cosh(kt)}$. Eventually, the Py thickness dependence of measured spin Hall angle $\theta_m$ could be described by

$$\theta_m=\theta_{sh}\frac{cosh(kt)-1}{cosh(kt)}. \qquad (1)$$

As shown in Fig. 3(a), the $\theta_m$ as a function of Py thickness is well fitted by Eq. (1) denoted by the red line. From fitting, we can deduce the intrinsic spin Hall angle $\theta_{sh}$ in Pt ($d = 6$ nm) to be $\sim 0.068 \pm 0.0025$ and the $k$ to be $\sim 0.78 \pm 0.09$ nm$^{-1}$. If we assume the spin diffusion length ($\lambda_S$) to be 5 nm in Py, the spin decoherence length ($\lambda_\phi$) is about 1.3 nm in Py. In addition, we have also estimated the spin pumping contributions to all Pt/Py($t$) devices.[39] We have found that the spin pumping contributions in our ST-FMR measurements are negligible and could not cause the Py thickness dependent spin Hall angle as shown in Fig. 3(a).

In order to get the spin diffusion length ($\lambda_S$) in Pt, another series of ST-FMR devices with the structures of Pt ($d = 3 - 14$ nm)/Py ($t = 4$ nm) has been also fabricated and characterized by taking the ratio of $S/A$. It must be noted that for the devices with Py thickness range from 4 to 7 nm, the values of measured spin Hall angle from both methods (by taking the ratio of $S/A$ and separate calibration, respectively) are almost same.[39] As shown in Fig. 3(b), the data of $\theta_m$ ($f = 8$ GHz) as a function of the Pt thickness match well with $\theta_m=\theta_\infty\left[1-\text{sech}\left(d/\lambda_S\right)\right]$,[10,11,31] where $\theta_m$ and $\theta_\infty$ represent the measured spin Hall angle at different Pt thickness ($d$) and a spin Hall angle at infinite Pt thickness, respectively.

From fitting, $\lambda_s$ in Pt is determined to be 1.5 nm.

As mentioned previously, there have been many reports on $\theta_{sh}$ in Pt by different groups using different techniques.[9-31] Here we summarize $\theta_{sh}$ from various reports and plot them together with our data in Fig. 3(c) as a function of the spin diffusion lengths ($\lambda_s$) in Pt. It must be noted that the $\theta_{sh}$ values measured by the ST-FMR technique represent a lower bound because of the possible spin scattering at the interface between the Py and Pt layers. From this plot, a clear correlation between $\theta_{sh}$ and $\lambda_s$ can be observed. The correlation is approximately inverse and can be reproduced by the product of $\theta_{sh}\lambda_s \sim 0.13$ nm (distribution range is denoted by the blue thick line).[42] The $\lambda_s$ in a material is generally proportional to its electrical conductivity ($\sigma$),[43] and we verify this relation in Fig. 3(d) by taking the data from the references in Fig. 3(c) and other reports.[31,36,44] All the data in Fig. 3(c-d) were measured at room temperature, except for those from Refs. 22 and 44 which were measured at 10 K, as denoted by small black stars.

A very wide range of $\sigma_{Pt}$ and thus $\lambda_s$ in previous studies indicates a possible reason for the significant disagreement in the reported $\theta_{sh}$. Moreover, it also opens up possibilities to tune the $\theta_{sh}$ by tuning $\sigma_{Pt}$ (and thus $\lambda_s$). The results shown in the inset of Fig. 3(d) demonstrate the possibility to engineer the spin Hall angle by tuning $\sigma_{Pt}$ via different *in-situ* annealing processes. Three series of samples with the same structure of Pt (6 nm)/Py (5 nm)/SiO$_2$ (3 nm) are fabricated, where the 6 nm Pt layer is firstly deposited on the wafers and then is *in-situ* annealed at different temperatures such as 250 and 350 $^o$C for 30 minutes. Subsequently, the Py and SiO$_2$ capping layers are deposited on top of Pt at room temperature. We find that the measured spin Hall angle in Pt increases as $\sigma_{Pt}$ decreases.

We have also investigated the temperature dependence of $\theta_m$ for Pt by ST-FMR measurements on a Pt (6 nm)/Py (5 nm) device at $f = 8$ GHz as shown in Fig. 4. The $\theta_m$ is determined by taking the ratio of $S/A$ and using $M_s$ of Py measured independently at corresponding temperatures. Interestingly, the $\theta_m$ exhibits a very small variation around 0.06 as temperature decreases from 300 to 13 K. Note that for other frequencies of 6 GHz and 10 GHz, the temperature dependence of $\theta_m$ shows the same behavior. On the other hand, the temperature dependence of the conductivity ($\sigma_{Pt}$) in Pt as shown in the inset of Fig. 4 is found to be significant. Since the intrinsic mechanism of the spin Hall angle implies a constant spin Hall conductivity ($\sigma_s$), the spin Hall angle ($\sigma_s/\sigma_{Pt}$) should decrease as temperature decreases, if an intrinsic mechanism dominates in our sample. However, from a relatively constant spin Hall angle in Fig. 4, we infer that the dominant mechanism for the spin Hall angle in Pt in our devices is not intrinsic.

In summary, we have studied the spin Hall angle of Pt in Pt/Py bilayers by Py thickness dependent ST-FMR measurements. By taking into account the previously neglected but important aspect of the spin diffusion in the Py layer, the intrinsic spin Hall angle in Pt ($d = 6$ nm) is determined to be $0.068 \pm 0.0025$ at room temperature. The $\lambda_s$ in Pt is measured to be 1.5 nm. A clear correlation between various $\theta_{sh}$ and $\lambda_s$ is established by the product of $\theta_{sh}\lambda_s \sim 0.13$ nm. Moreover, we find that the spin Hall angle is almost independent of temperature in the range of 13 to 300 K.

This research is supported by the National Research Foundation, Prime Minister's Office, Singapore under its Competitive Research Programme (CRP Award No. NRF-CRP 4-2008-06 and NRF-CRP12-2013-01).

**Figure captions**

FIG. 1. (a) The schematic diagram of the ST-FMR measurement configuration. (b) The Pt/Py microstrip with the charge current $J_c$, spin current $J_s$, RF Oersted field ($H_{rf}$). (c) The measured ST-FMR signals (open symbols) on Pt (6 nm)/Py (5 nm) device with fits (solid lines). (d) Resonance frequency $f$ against the resonant field $H_0$ for Pt (6 nm)/Py ($t$ nm) devices with fits (solid lines).

FIG. 2. (a) The normalized values of $S$ (symmetric component) and $A$ (antisymmetric component) extracted from the ST-FMR fitting as a function of magnetic field angles ($\theta_H$) for a Pt (6 nm)/Py (5 nm) device at $f = 8$ GHz and the corresponding fits (solid lines). The normalized $S$ values have an offset of -0.3 for clarity. (b) The field angle ($\theta_H$) dependence of $\theta_m$ in Pt for Pt (6 nm)/Py (5 nm) device at 8 GHz and 300 K.

FIG. 3. (a) $\theta_m$ for Pt (6 nm)/Py ($t$) devices with different Py thicknesses from ST-FMR measurements at 8 GHz and 300 K. The dashed blue lines ($a$, $b$, and $c$) represent three cases of zero thickness limit of the Py layer. The solid red line shows the fit using Eq. (1). (b) $\theta_m$ (blue squares) on Pt ($d$)/Py (4 nm) devices as a function of Pt thickness at 300 K, and a fit (solid red line). (c) $\theta_{sh}$ as a function of $\lambda_S$ in Pt from various reports denoted by the reference numbers. The STFMR, SP, LSV, and SMR represent spin torque ferromagnetic resonance, spin pumping, nonlocal measurement in lateral spin valves, and spin Hall magnetoresistance method, respectively. The blue thick curve shows a correlation $\theta_{sh}\lambda_S \sim$ 0.13 nm. (d) $\lambda_S$ in Pt as a function of its electrical conductivity ($\sigma_{Pt}$). The inset shows the measured spin Hall angles in various Pt films with different conductivities. The dashed lines

are guides to the eye.

FIG. 4. The temperature dependence of $\theta_{\mathrm{m}}$ in Pt determined by ST-FMR measurements. The inset shows the temperature dependence of the Pt electrical conductivity ($\sigma_{\mathrm{Pt}}$).

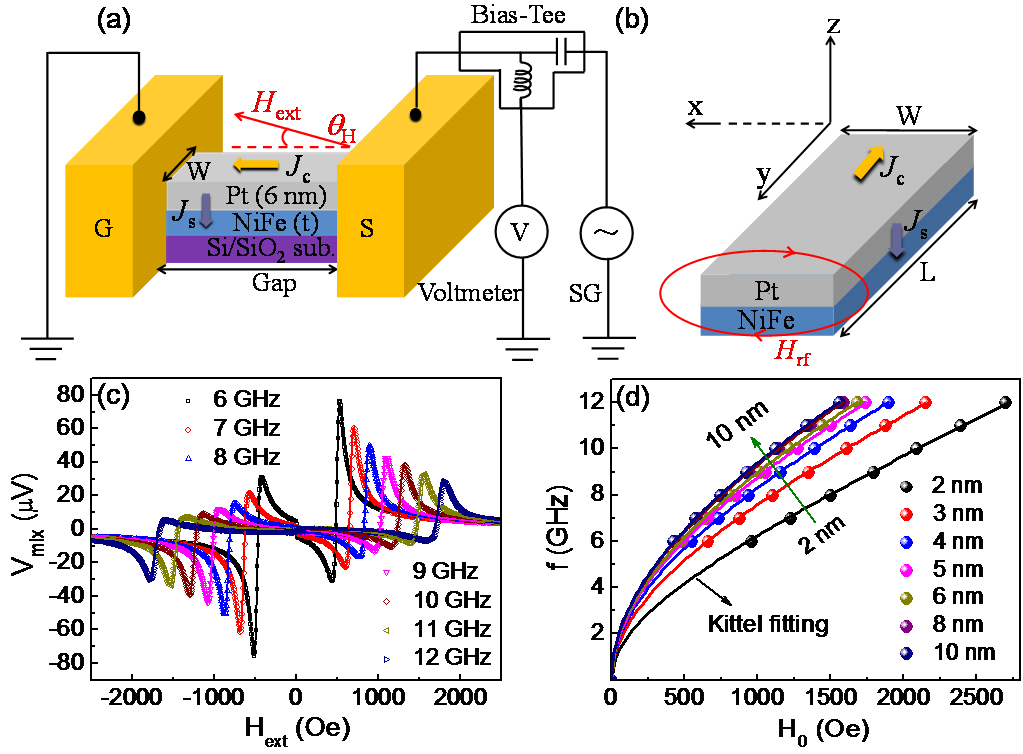

FIG. 1

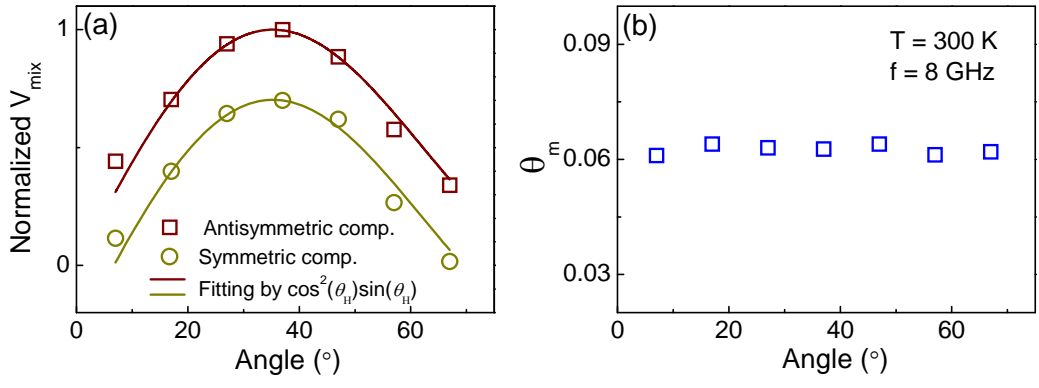

FIG. 2

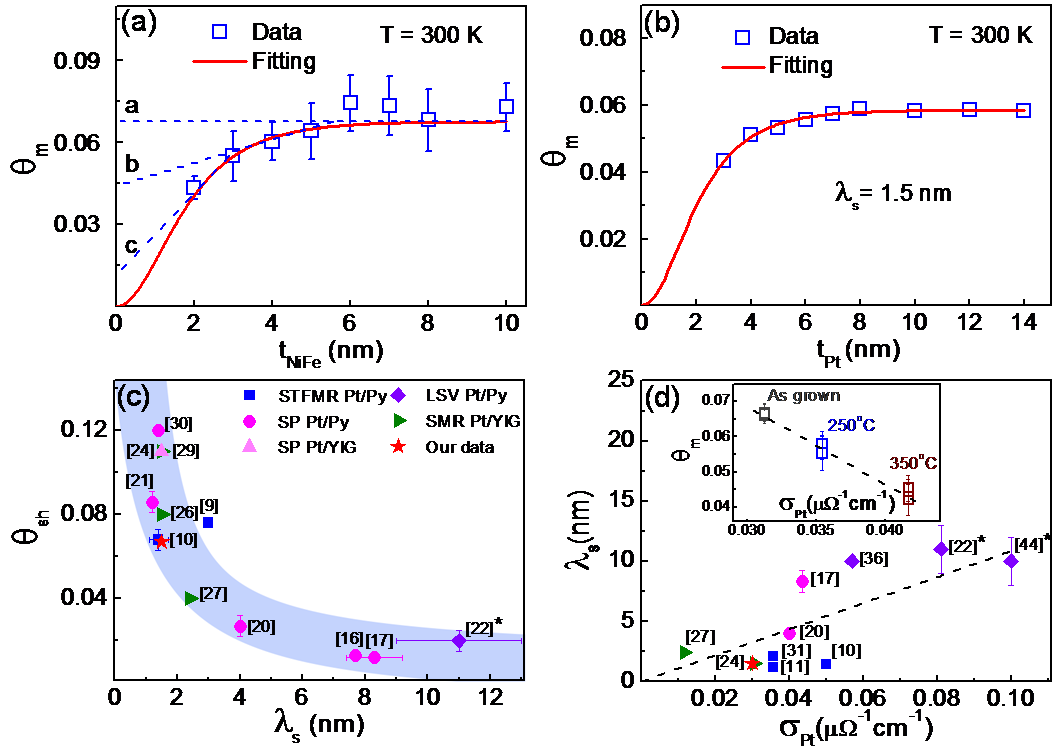

FIG. 3

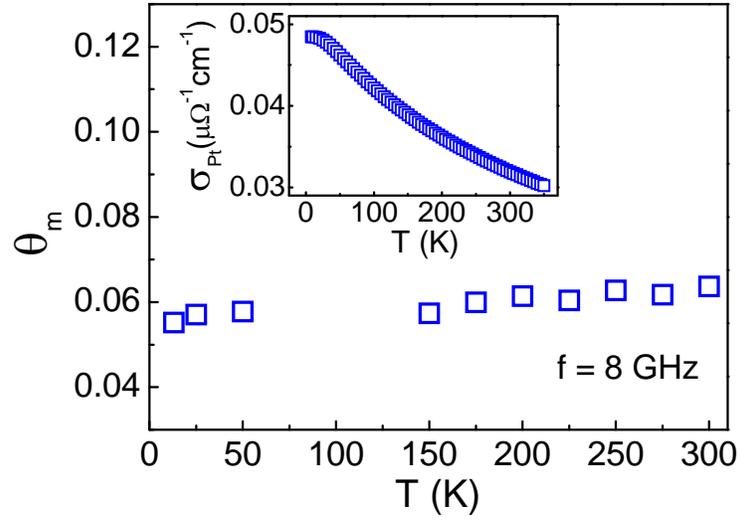

FIG. 4